\documentclass[aps,twocolumn,superscriptaddress,showkeys,showpacs]{revtex4}
\usepackage{amssymb}
\usepackage{amsfonts}
\usepackage{amsmath}
\usepackage{amsthm}
\usepackage{epsfig}
\usepackage{color} 
\usepackage{subfigure}
\usepackage{bbm}

\newcommand{\eps}{\varepsilon}

\def\mean#1{\langle#1\rangle}

\begin{document}
\title{Noise-induced transitions in a double-well excitable oscillator}

\author{Vladimir V. Semenov}
\email{semenov.v.v.ssu@gmail.com}
\affiliation{Department of Physics, Saratov State University, Astrakhanskaya str., 83, 410012, Saratov, Russia}

\date{\today}

\begin{abstract}

The model of a double-well oscillator with nonlinear dissipation is studied. The self-sustained oscillation regime and the excitable one are described. The first regime consists in the coexistence of two stable limit cycles in the phase space, which correspond to self-sustained oscillations of the point mass in either potential well. The self-sustained oscillations do not occur in a noise-free system in the excitable regime, but appropriate conditions for coherence resonance in either potential well can be achieved. The stochastic dynamics in both two regimes is researched by using numerical simulation and electronic circuit implementation of the considered system. Multiple qualitative changes of the probability density function (PDF) caused by noise intensity varying are explained by using the phase space structure of the deterministic system. 

\end{abstract}
\pacs{ 02.30.Ks, 05.10.-a, 05.40.-a, 84.30.-r}
\keywords{bistability, double-well oscillator, noise, noise-induced transitions, coherence resonance}
\maketitle

\section{Introduction} 
Bistable systems are of frequent occurrence in physics \cite{gibbs1985,hanggi1990, risken1996}, chemistry \cite{kramers1940,schloegl1972}, biology \cite{goldbetter1997,guidi1998,ozbudak2004,shilnikov2005,shilnikov2012,smits2006,benzi2010}, ecology \cite{may1977,guttal2007}, geophysics \cite{benzi1981,benzi1982,nicolis1982}. Stochastic bistable systems attract an interest because of a wide range of noise-induced phenomena in such systems: stochastic resonance \cite{gammaitoni1998,anishchenko1999}, doubly stochastic coherence \cite{zaikin2003}, noise-induced chaos \cite{schimansky1993} and synchronization \cite{neiman1994,shulgin1995}, suppression of the chaotic dynamics by random perturbations \cite{gassmann1997}. The noise-driven bistable systems can also exhibit noise-induced transitions whereby the stationary PDF changes its structural shape, e.g.  number of extrema, when noise intensity varies \cite{horsthemke1984}. Such transitions  may occur both with multiplicative noise as in the original Horsthemke-Lefever scenario \cite{horsthemke1984}, and with additive noise (see e.g. \cite{schimansky1985,zakharova2010}). Noise-induced transitions were observed in excitable systems ranging from a single excitable neuron \cite{tanabe2001,kromer2014,neiman2007} to coupled excitable elements and media \cite{ullner2003,zaks2005}. In many cases noise-induced transitions are not true bifurcations \cite{toral2011},  rather they underlie qualitative changes of the stochastic dynamics. Noise-induced {\it phase} transitions were studied in spatially distributed systems perturbed by multiplicative noise \cite{van1994} and were shown to exist for the case of additive noise \cite{zaikin1999}. 

A classical example of the stochastic bistable system is the Kramers oscillator describing Brownian motion in a double-well potential \cite{hanggi1990,kramers1940,freund2003},
\begin{equation}
\label{kramers}
\dot{y}=v, \quad \dot{v} = -\gamma v - \frac{dU(y)}{dy} + \sqrt{2\gamma D}\,n(t),
\end{equation}
where $\gamma$ is the drag coefficient, $U(y)$ is a potential function, $n(t)$ is Gaussian white noise, and $D$ is the noise intensity. If $\gamma$ is a constant parameter, then the noise-free system exhibits the simplest kind of the bistability: the coexistence of two stable fixed points in the phase space. In that case the two-dimensional stationary PDF of the stochastic system (\ref{kramers}) possesses two maxima corresponding to potential wells, which are separated by a saddle point of the potential. Position and number of peaks are invariant with respect to increase of the noise intensity, and a structure of the PDF does not depend on the noise intensity. The dynamics of the double-well potential systems like system (\ref{kramers}) can be essentially different in the presence of dissipation depending nonlinearly on the system state. In particular, the bistable model offered in \cite{semenov2016}, which can be reduced to the Eq.~(\ref{kramers}) with the nonlinear drag function $\gamma=\gamma(y,\dot{y})$ and the double-well potential $U(y)$, demonstrates multiple noise-induced transitions and the non-monotonic dependence of the Rice frequency on the noise intensity. 

Initially, the introduced in \cite{semenov2016} oscillator was considered in the simplest regime. Certain peculiarities of the considered system were excluded from consideration. In particular, the phase space structure corresponding to chosen parameter values did not allow the system to exhibit properties of excitable oscillators. The system dynamics is not limited to the regime presented in \cite{semenov2016}. The self-oscillatory regime and the excitable one, which is associated with the effect of coherence resonance, also are realized in the system. However, the stochastic dynamics of the system in these two regimes is essentially complicated. Correct analysis and interpretation of noise-induced transitions in the self-oscillatory and excitable regimes can be carried out on base of the results obtained in \cite{semenov2016}. In the present paper the oscillator proposed in \cite{semenov2016} is explored in the excitable regime and in the regime of the self-sustained oscillations. Multiple noise-induced transitions due to additive noise are explained by the phase space structure of the deterministic system. Numerical simulation of the system under study is combined with real experiments on the example of the corresponding electronic analog model. 

\section{Model and Methods}
The system under study is described by the following equations:

\begin{equation}
\label{system}
\left\lbrace
\begin{array}{l}
\varepsilon \dot{x} = -y-c_{1}x+c_{3}x^3-c_{5}x^5-\sqrt{2D}n(t), \\
\dot{y}=x+ay-by^3, \\
\end{array}
\right.
\end{equation}
 where $x$ and $y$ are the dimensionless variables, $\dot{x}=\dfrac{dx}{dt}$, $\dot{y}=\dfrac{dy}{dt}$, $t$ is the dimensionless time, the parameters $c_{1,3,5},a,b >0$ define nonlinearity, the parameter $\eps$ sets separation of slow and fast oscillations in the system, $n(t)$ is normalized Gaussian white noise: $\mean{n(t)}=0$, $\mean{n(t)n(t+\tau)}=\delta(\tau)$, and $D$ is the noise intensity. Eqs. (\ref{system}) can be written in the "coordinate-velocity" form with the dynamical variables $y$, $v\equiv \dot{y}$ and $x=v-ay+by^3$:
\begin{equation}
\label{system2}
\left\lbrace
\begin{array}{l}
\dot{y}=v, \\
\eps \dot{v} =-y-c_1 (v-ay+by^3)+c_3 (v-ay+by^3)^3-\\ 
c_5 (v-ay+by^3)^5+ \eps v ( a -3 by^2)-\sqrt{2D}n(t).
\end{array}
\right.
\end{equation}
In the oscillatory form (\ref{system2}) becomes,
\begin{equation}
\label{system3}
\ddot{y}+q_{1}(y,\dot{y})\dot{y}+\frac{1}{\varepsilon}q_{2}(y)=-\sqrt{2 D}n(t),
\end{equation}
and describes motion in a potential field, where $\dfrac{1}{\varepsilon}q_{2} (y)=\dfrac{dU(y)}{dy}$ defines the form of the potential, $U(y)$, and $q_{1}(y, \dot{y})$ is the nonlinear dissipation,
\begin{equation}
\label{q1q2}
\left.
\begin{array}{l}
q_{1}(y,\dot{y})=-a+3by^{2}+ \dfrac{1}{\varepsilon}(c_{1}-\\
c_{3} \sum\limits_{n=1}^{3} \frac{3!}{n!(3-n)!} \dot{y}^{n-1} (by^{2}-a)^{3-n}y^{3-n}+ \\
c_{5} \sum\limits_{n=1}^{5} \frac{5!}{n!(5-n)!} \dot{y}^{n-1} (by^{2}-a)^{5-n}y^{5-n}), \\
 \\
q_{2}(y)= y+c_{1}(by^{2} -a)y-\\
\\
c_{3}(by^{2} -a)^{3} y^{3} +c_{5}(by^{2} -a)^{5} y^{5}. \\
\end{array}
\right.
\end{equation}

Further consideration of the system will be carried out in the variables ($y,v=\dot{y}$) (the systems (\ref{system2}) and (\ref{system3})). In the following the parameters are set to $\varepsilon=0.01$, $b=50$, $c_{1}=1$, $c_{3}=9$, $c_{5}=22$. In this case increasing of the parameter $a$ from zero gives rise to the following bifurcational changes in the phase space of the deterministic system. Initially, there exists one stable fixed point in the origin. When the parameter $a$ reaches the value $a=1/c_{1}=1$, a supercritical pitchfork bifurcation occurs: the stable fixed point in the origin becomes unstable (saddle point) and two stable fixed points appear at the left and at the right of the origin. Further increasing the parameter $a$ leads to the loss of stability of two side fixed points, and two stable limit cycles appear at the vicinity of the unstable points of equilibtium. It is the Hopf scenario of soft self-sustained oscillation excitation, which is realized at the same moment in two points at $a\approx2.494$. As will be shown below, the system exhibits the excitable behavior before the pair Hopf bifurcation. The present paper, as distinct from \cite{semenov2016}, is focused on the studying of system (\ref{system2}) near the Hopf bifurcation: in the excitable regime at $a=2.4$ and in the self-oscillatory regime at $a=3$. 

The explored system was studied by means of analog and numerical simulations. Numerical modelling of the considered system was carried out by integration of Eq. (\ref{system2}) using the Heun method \cite{manella2002} with the time step $\Delta t=0.0001$. The total integration time is $t_{int}=10^{6}$. Experimental electronic setup is described in the paper \cite{semenov2016} in details. For this reason description and illustration of the circuit diagram are absent in the present paper. Time series of the experimental facility were recorded from the corresponding outputs by using an acquisition board (National Instruments NI-PCI 6133). All signals were digitized at the sampling frequency of 50 kHz. 150~s long realizations were used for offline time series analysis. The noise generator G2-59 was used to produce broadband Gaussian noise, whose spectral density was almost constant in the frequency range 0  -- 100 kHz, therefore noise can be assumed to be white in this frequency range.

\section{Regime of self-sustained oscillations}
\subsection{Noise-free system}

The phase space of the deterministic (with $D=0$) system (\ref{system2}) in the self-sustained oscillations regime is presented in Fig.~\ref{fig1}(a). All trajectories are attracted to stable limit cycles in the left and right half-spaces. Basins of attraction of the stable limit cycles are separated by the separatrix (the blue dotted line in Fig.~\ref{fig1}(a)) of the saddle at the origin. The self-sustained oscillations represent the fast-slow dynamics like in the FitzHugh-Nagumo model \cite{izhikevich2006}. It includes slow motions along the nullcline $\dot{v}=0$ and the fast ones, when the phase point falls down from the nullcline. Investigation of the system in the oscillatory form (Eqs.~(\ref{system3}) and (\ref{q1q2})) allows us to reveal nature of this regime. In the vicinity of the left and right equilibrium points the dissipation, $q_{1}(y,v)$, is negative [Fig.~\ref{fig1}(b)]. It denotes energy pumping, which leads to the instability of both the left and right equilibria, and the self-sustained oscillations excitation occurs. Energy balance between dissipation and pumping during the period of the self-sustained oscillations is organized after short transient time. The shape of the self-sustained oscillations is determined by the nonlinear dissipation, $q_{1}(y,v)$, and by the potential function, $U(y)$, which can be calculated as an integral of the function $q_{2}(y)$ divided by $\varepsilon$. The potential function corresponding to the bistable self-sustained oscillation regime has a double-well form [Fig.~\ref{fig1}(c)]. In such a way, this dynamics can be considered as two coexisting self-sustained oscillation regimes (the regimes 1 and 2 are schematically marked in Fig.~\ref{fig1}(c)) of active Brownian particle in a double-well potential field. Fragments of the $y(t)$ and $v(t)$ time realizations in the regimes 1 and 2 are illustrated in Fig.~\ref{fig1}(d). 

\subsection{Noise-induced transitions}
\begin{figure}[t!]
\centering
\includegraphics[width=0.5\textwidth]{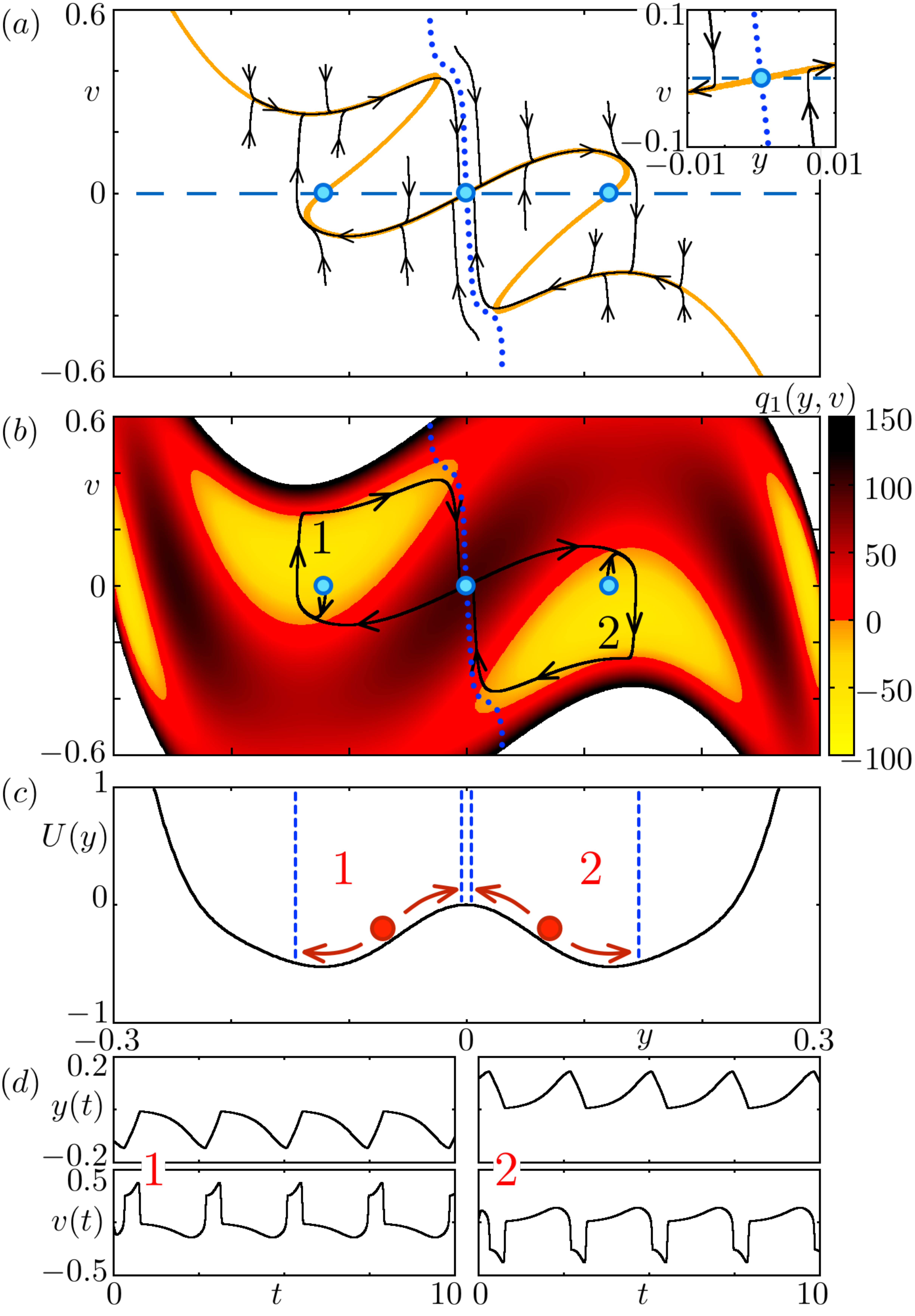} 
\caption{Noise-free system (\ref{system2}) in the self-sustained oscillation regime in numerical simulation. (a) Phase space. Equilibrium points are shown by blue circles; the blue dashed line indicates the nullcline $\dot{y} = 0$; the orange solid line shows the nullcline $\dot{v}= 0$; the separatrix of the saddle at the origin is shown by the blue dotted line. Phase trajectories started from various initial conditions are shown by black arrowed lines. (b) Dependence of the dissipation, $q_{1}(y,v)$, on the system state (Eq.~\ref{q1q2}). Stable limit cycles corresponding to the self-sustained oscillation regimes 1 and 2 are marked by black closed curves. (c) Potential function, $U(y)$, corresponding to $q_{2}(y)$ (Eq.~\ref{q1q2}). Self-sustained oscillations in either potential well 1 and 2 are schematically shown. (d) Time traces of state variables in two coexisting self-sustained oscillatory regimes 1 and 2. Parameters are: $\varepsilon=0.01$, $c_{1}=1$, $c_{3}=9$, $c_{5}=22$, $a=3$, $b=50$, $D=0$.}
\label{fig1}
\end{figure}

\begin{figure}[t!]
\centering
\includegraphics[width=0.5\textwidth]{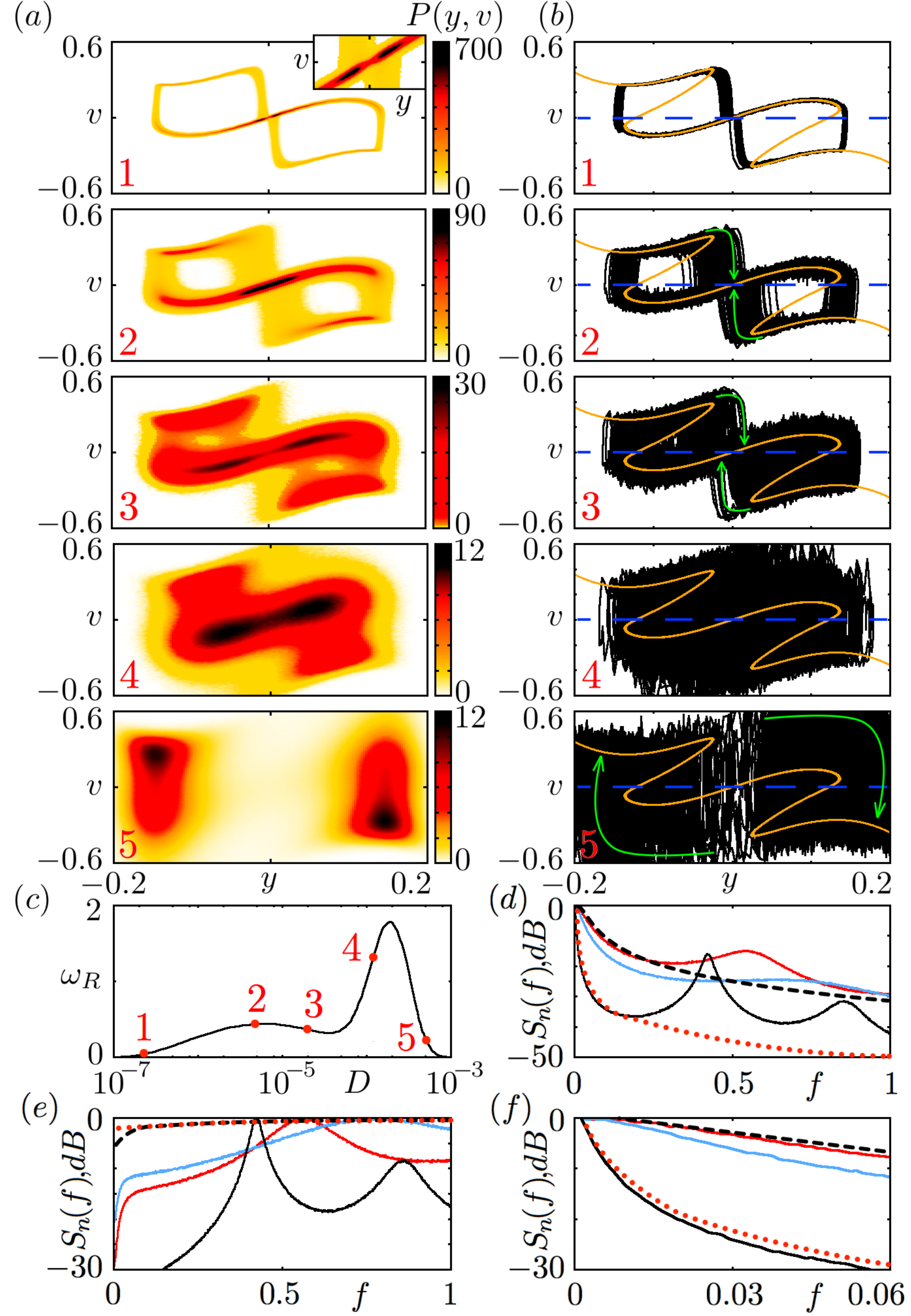} 
\caption{Noise-induced transitions in the system (\ref{system2}) in the self-sustained oscillation regime (numerical simulations). The evolution of the PDF when the noise intensity increases (a) and the corresponding fragments of the phase trajectory (b): 1, $D=2.5\times 10^{-7}$; 2, $D=5\times 10^{-6}$; 3, $D=2\times 10^{-5}$; 4, $D=1.2\times10^{-4}$; 5, $D=5\times 10^{-4}$. On all panels: the blue dashed line indicates the nullcline $\dot{y} = 0$; the orange solid line shows the nullcline $\dot{v} = 0$. 
(c) Rice frequency, $\omega_{R}$, vs noise intensity, $D$. 
(d),(e) Evolution of the normalized power spectrum of the $y(t)$-oscillations (d) and $v(t)$-oscillations (e), $S_{n}(f)=S(f)/S_{max}(f)$, when the noise intensity increases: $D=2.5\times 10^{-7}$ (the black solid curve), $D=5\times 10^{-6}$ (the red solid curve), $D=2\times 10^{-5}$ (the blue solid curve), $D=1.2\times10^{-4}$ (the black dashed curve), $D=5\times 10^{-4}$ (the red dotted curve). (f) Normalized power spectra of the $y(t)$-oscillations (the panel (d)) illustrated in more details in the low frequency range.
Other parameters are: $\varepsilon=0.01$, $c_{1}=1$, $c_{3}=9$, $c_{5}=22$, $a=3$, $b=50$.}
\label{fig2}
\end{figure}

Numerical simulation of the system (\ref{system2}) has shown that noise strength is the true control parameter of the system. For weak noise the system exhibits the noisy bistable dynamics with typical hopping between two self-sustained oscillation regimes [Fig.~\ref{fig2}(b1)]. The corresponding PDF consists of two separated closed craters [Fig.~\ref{fig2}(a1)]. Increase of the noise intensity leads to qualitative change in the PDF: two local peaks of the closed craters merge into one peak in the origin [Fig.~\ref{fig2}(a2)]. This transition is caused by two effects. The first reason consists in the smearing of two peaks, and the second one is that the phase point more frequently hits the vicinity of the saddle point of equilibrium, where dissipation increases [Fig.~\ref{fig1}(b)] and the phase point becomes slowed down (see green arrowed lines in Fig.~\ref{fig2}(b2)). Further increasing the noise intensity results to destruction of the closed craters in sectors corresponding to the fast phase of the self-sustained oscillations. Partial destruction of the closed crater in the PDF is typical for anharmonical self-sustained oscillators and is associated with the existence of so-called bifurcational interval at the Hopf bifurcation \cite{ebeling1986,fronzoni1987,anishchenko2014}. Phase trajectories can now overcome the separatrix towards another closed crater, rather than fall onto the origin (see green arrowed lines in Fig.~\ref{fig2}(b3)). As a result, the central peak of the PDF is split up into two peaks [Fig.~\ref{fig2}(a3)]. If the noise intensity goes up to increase, then phase point drift becomes more stronger and the peaks of the PDF are smeared and the closed craters destruction finishes [Fig.~\ref{fig2}(a4,b4)]. The tendency to central peak forming in the origin is seen. It happens because the phase point frequently reaches the vicinity of the saddle point in the origin from different areas of the phase space and then becomes slowed down. However, it is insufficient for final shaping of the central peak of the PDF in the origin. For larger values of the noise intensity vertical moving of the phase point results in its more intensive left-right shifting (see green arrowed lines in Fig.~\ref{fig2}(b5)). Then the phase point is slowed down on attractive branches of the nullcline $\dot{v}=0$, and two peaks in the PDF are formed [Fig.~\ref{fig2}(a5)]. It  results in conventional stochastic hopping between two metastable states with the double-peaked PDF [Fig.~\ref{fig2}(a5)].  Signs of the self-sustained oscillatory regime finally disappear. 
%

%

The described noise-induced transitions result in the nonmonotonic dependence of the Rice frequency on the noise intensity with two maxima [Fig.~\ref{fig2}(c)]. The Rice frequency is the rate of zero-crossings by the oscillator's coordinate with positive velocity, $\omega_R = 2\pi \int_0^\infty v P(y=0,v) dv$, and characterizes the mean frequency of bistable oscillators \cite{callenbach2002,freund2003}. Merge of the peaks in the PDF into one central peak or striving for central maximum formation logically leads to the Rice frequency increasing. On the contrary, division of the central peak with the noise-intensity growth leads to the Rice frequency decreasing or deceleration of its increasing. In a case of extremely large noise the PDF peaks become more devious, and then the Rice frequency tends to zero. 

Noise-induced phenomena in the considered system are reflected in the power spectra of the oscillations $y(t)$ [Fig.~\ref{fig2}(d)] and $v(t)$ [Fig.~\ref{fig2}(e)]. In the presence of weak noise the power spectrum has a spectral peak corresponding to the self-sustained oscillations in the potential wells. This local peak gets smeared and then finally disappears when the noise intensity growths. Destruction of the closed craters in the PDF and disappearance of the local peak in the power spectrum occur simultaneously. Therefore one can say about noise-induced destruction of the self-sustained oscillations in the system (\ref{system2}). In the range of low frequencies the power spectrum of the $y(t)$ oscillations has a Lorentzian shape with a width nonmonotonically changing with the noise intensity growth [Fig.~\ref{fig2}(d,f)]. The following trend is revealed in the system (\ref{system2}): the bigger is the Rice frequency, the wider is the Lorentzian in the power spectrum. The existence of the Lorentzian in the power spectrum is typical for the double-well oscillators similar to Eq.  (\ref{kramers}) with constant dissipation. The dynamics of the system (\ref{kramers}) can be described as a stochastic telegraph process. The power spectrum of such process has a Lorentzian shape with a width being proportional to mean rate of switching events. Despite the similar transformation of the power spectrum of the $y(t)$-oscillations in the system (\ref{system2}), the full analogy is incorrect. The dynamics of the system (\ref{system2}) has more complex character and cannot be reduced to the random telegraph process. The PDF including a local maximum in the origin (in the saddle point) does not correspond to transitions between two attractors.
 
\begin{figure}[t]
\centering
\includegraphics[width=0.5\textwidth]{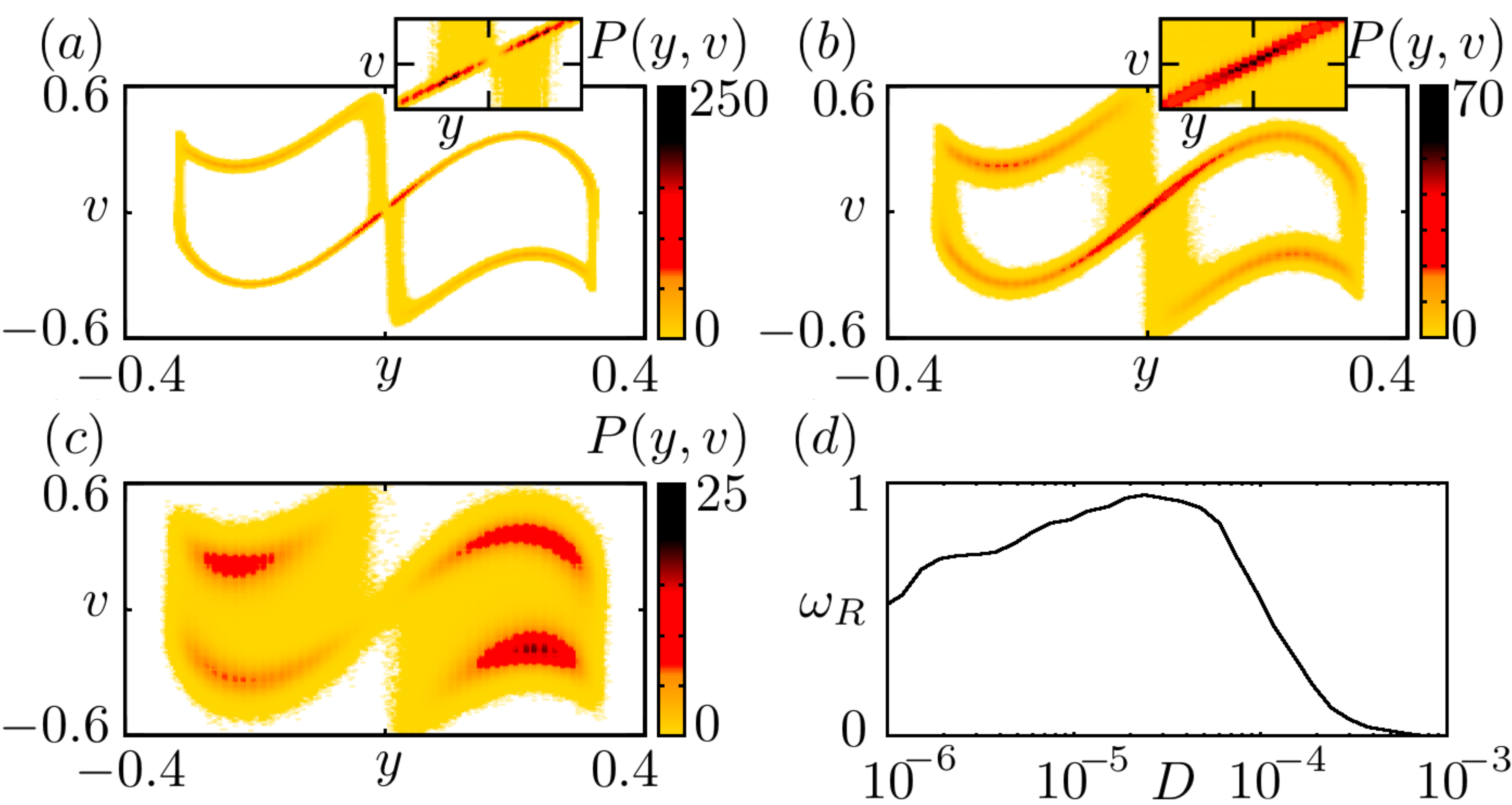} 
\caption{Self-sustained oscillatory regime in analog experiment. (a)-(c) Evolution of the PDF when the noise intensity increases: (a) $D=10^{-6}$; (b) $D=2.4\times10^{-5}$; (c) $D=6\times10^{-6}$. (d) Rice frequency, $\omega_{R}$, vs noise intensity, $D$. Parameters are: 
$\varepsilon=0.01$, $c_{1}=1$, $c_{3}=9$, $c_{5}=22$, $a=2.4$, $b=50$.}
\label{fig3}
\end{figure}  
%

Despite the direct correspondence between experimental setup equations and Eqs. (\ref{system2}), the behavior of the analog circuit had somewhat different character.  This is due to the fact that the experimental facility equations were derived using standard approximations on operation amplifiers, which are common in electronics (a model of the ideal operation amplifier was used). Inaccuracies of parameter measurements also took a part. The experimental facility had its own inevitable presenting dynamical noise, which was sufficient for hopping between two coexisting self-sustained oscillatory regimes without external influence. The electronic model also turned out very sensitive to the features of the external noise generator. External noise signal used in analog experiments had very small but non-zero mean value, which unpredictable drifted during the experiment. As a result, the experimentally obtained PDF are a little bit unsymmetrical, and the dependence of the Rice frequency on the noise intensity was more rough as compared to dependence $\omega_{R}(D)$ obtained in numerical modelling. Without external noise, or in the presence of weak noise the experimental device demonstrates two separated closed craters in the PDF [Fig.~\ref{fig3}(a)]. Two local maxima of each crater merge into one peak in the origin when the noise intensity growths [Fig.~\ref{fig3}(b)]. Further increasing of the noise intensity gives rise to splitting of the peak in the origin and to partial destruction of the closed craters [Fig.~\ref{fig3}(c)]. Formed PDF peaks promptly move away from the origin with further noise intensity growth. Only two qualitative changes in the PDF are observed in analog experiment: formation and splitting of the peak in the origin. Resulting dependence of the Rice frequency on the noise intensity has one maximum. The experimental dependence $\omega_{R}(D)$ [Fig.~\ref{fig3}(d)] was normalized with taking into account the difference between time scales of the mathematical model (\ref{system2}) and the experimental setup.

\section{Excitable regime}
\subsection{Pair coherence resonance}
\begin{figure}[t!]
\centering
\includegraphics[width=0.5\textwidth]{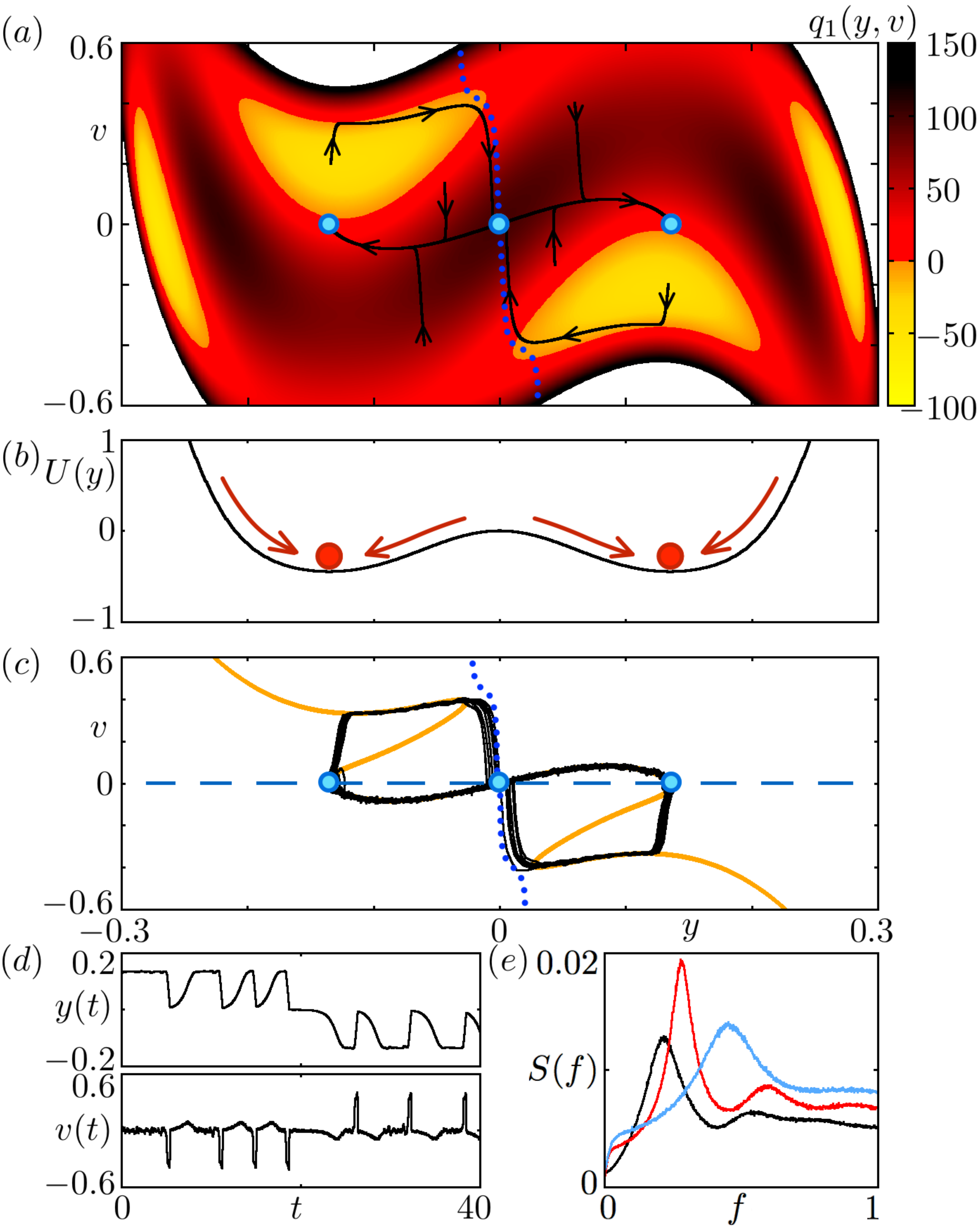} 
\caption{Excitable regime in numerical simulations (system (\ref{system2})). (a) Dependence of the dissipation, $q_{1}(y,v)$, on the system state (Eq.~\ref{q1q2}). Trajectories in a noise-free system ($D=0$) started from various initial conditions and coming to stable fixed points are marked by black arrowed lines. (b) Potential function, $U(y)$, corresponding to $q_{2}(y)$ (Eq.~\ref{q1q2}). Stability in either well is schematically shown. (c) Phase space of the system (\ref{system2}). Equilibrium points are shown by blue circles; the blue dashed line indicates the nullcline $\dot{y} = 0$; the orange solid line shows the nullcline $\dot{v}= 0$; the separatrix of the saddle at the origin is shown by the blue dotted line. Noise-induced oscillations  ($D=2.5\times 10^{-7}$) are shown by the black trajectory. (d) Time traces of state variables corresponding to the fragment (c). (e) Power spectra of the $v(t)$ noise-induced oscillations for various values of the noise intensity: $D=2\times 10^{-7}$ (the black curve);  $D=10^{-6}$ (the red curve); $D=10^{-5}$ (the blue curve). Parameters are: 
$\varepsilon=0.01$, $c_{1}=1$, $c_{3}=9$, $c_{5}=22$, $a=2.4$, $b=50$.}
\label{fig4}
\end{figure}  
%
The system's (\ref{system2}) phase space structure allows to create appropriate conditions for the effect of coherence resonance like in the FitzHugh-Nagumo model \cite{pikovsky1997,lindner1999,lindner2004} both in the left and right half-spaces. Next the parameter $a$ of the system (\ref{system2}) is fixed as being $a=2.4$, which is slightly lower than the Hopf bifurcation value. The chosen parameter value corresponds to the coexistence of two stable fixed points. Dissipation is positive in the vicinity of the stable equilibria [Fig.~\ref{fig4}(a)]. In the presence of noise the random force occasionally kicks the phase point out of the vicinity of the stable equilibria towards the region of negative friction. Phase point drift can be amplified in the areas of negative friction, and then the phase point trajectory forms a loop. The potential function, $U(y)$, remains to be double-well as well as in the self-sustained oscillatory regime [Fig.~\ref{fig4}(b)]. Fluctuations are responsible for noise-induced oscillations within either potential well and for transitions between two wells. Consequently, the noise-induced oscillations in the phase space include motions along two loops and transitions between them [Fig.~\ref{fig4}(c)]. 
The oscillations along the nullcline loops in the phase plane are manifested as spikes in the $v(t)$ time-realizations [Fig.~\ref{fig4}(d)]. There is an optimal noise intensity corresponding to the most regular spiking. Enhancement of the correlation is more evident on the power spectrum of the $v(t)$-oscillations [Fig.~\ref{fig4}(e)]. The main spectral peak initially increases and becomes more narrow with the noise intensity growth, but then it decreases and fades out. That transformation of the power spectra is typical for the classical coherence resonance in excitable systems.
\subsection{Noise-induced transitions}
\begin{figure}[t!]
\centering
\includegraphics[width=0.5\textwidth]{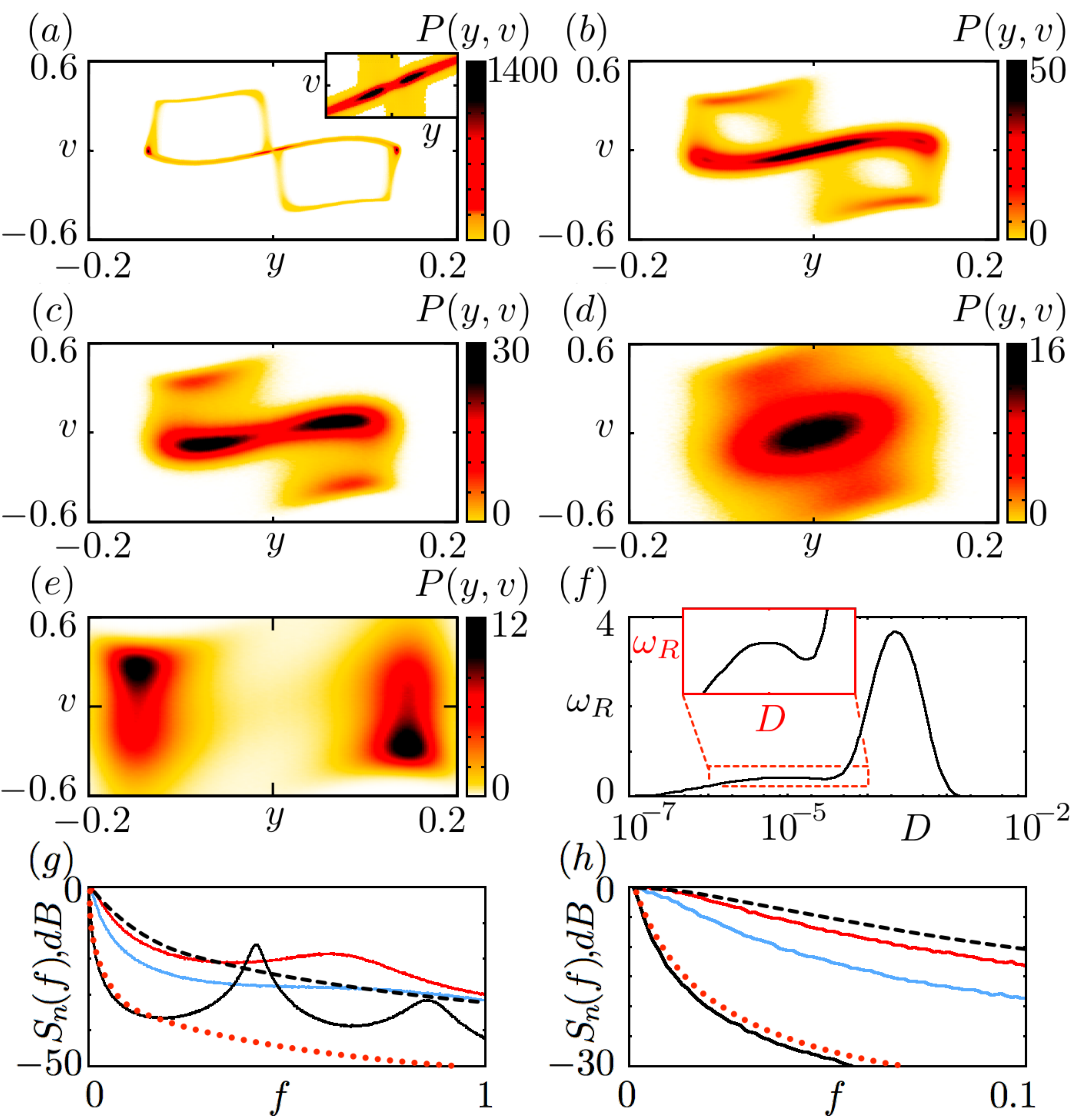} 
\caption{Excitable regime in numerical simulations (system (\ref{system2})). (a)-(e) Evolution of the PDF when the noise intensity increases: (a) $D=2.5\times 10^{-7}$; (b) $D=10^{-5}$; (c) $D=3\times 10^{-5}$; (d) $D=2\times 10^{-4}$; (e) $D=5\times 10^{-4}$. (f) Rice frequency, $\omega_{R}$, vs noise intensity, D. (g) Evolution of the normalized power spectrum of the $y(t)$-oscillations, $S_{n}(f)=S(f)/S_{max}(f)$, when the noise intensity increases: $D=2.5\times 10^{-7}$ (the black solid curve), $D=10^{-5}$ (the red solid curve), $D=3\times 10^{-5}$ (the blue solid curve), $D=2\times10^{-4}$ (the black dashed curve), $D=5\times 10^{-4}$ (the red dotted curve). (h) Normalized power spectra of the $y(t)$-oscillations illustrated in more details in the low frequency range. Parameters are: $\varepsilon=0.01$, $c_{1}=1$, $c_{3}=9$, $c_{5}=22$, $a=2.4$, $b=50$.
}
\label{fig5}
\end{figure}  
%
The stochastic dynamics of the system (\ref{system2}) in the excitable regime strongly resembles described above noise-induced transitions corresponding to the self-sustained oscillatory regime. In the presence of weak noise the effect of coherence resonance gives rise to closed craters formation in the PDF, which corresponds to the noise-induced motions along the nullcline loops [Fig.~\ref{fig5}(a)]. Either closed crater has two peaks. The first peak corresponds to motions of the phase point in the vicinity of the stable equilibria, the second one is a result of the slowed motions in the vicinity of the saddle point in the origin (it is a region of large dissipation). Two peaks in the PDF situated close to the origin merge into the central peak [Fig.~\ref{fig5}(b)] with the noise intensity growth. Further noise-intensity increasing results in division of the central peak into two peaks and closed craters destruction [Fig.~\ref{fig5}(c)]. Noise induced changes in the PDF have the same reasons in the self-sustained oscillatory regime and in the excitable one. The main qualitative difference arises when two formed peaks [Fig.~\ref{fig5}(c)] merge into one central peak again [Fig.~\ref{fig5}(d)]. As in the self-sustained oscillatory regime, it happens because the phase point reaches the saddle equilibrium vicinity from different areas of the phase space and then becomes slowed down. However, in contrast to the self-sustained oscillatory regime, the central peak is finally forming. This peak becomes divided again [Fig.~\ref{fig5}(e)] when the noise intensity growth continues. A consequence of the described transformations in the PDF is the nontrivial dependence of the Rice frequency on the noise-intensity with two-maxima [Fig.~\ref{fig5}(f)], which is similar to the presented above curve [Fig.~\ref{fig2}(c)] corresponding to the self-sustained oscillation regime. 

Evolution in the $y(t)$-oscillation power spectrum is similar to spectral transformation exhibited in the regime of the self-sustained oscillations [Fig.~\ref{fig5}(g)]. Width of the Lorentzian changes nonmonotonically when the noise intensity increases [Fig.~\ref{fig5}(h)], as well as the spectral peak caused by the effect of coherence resonance becomes smeared and disappears. 

\begin{figure}[t]
\centering
\includegraphics[width=0.5\textwidth]{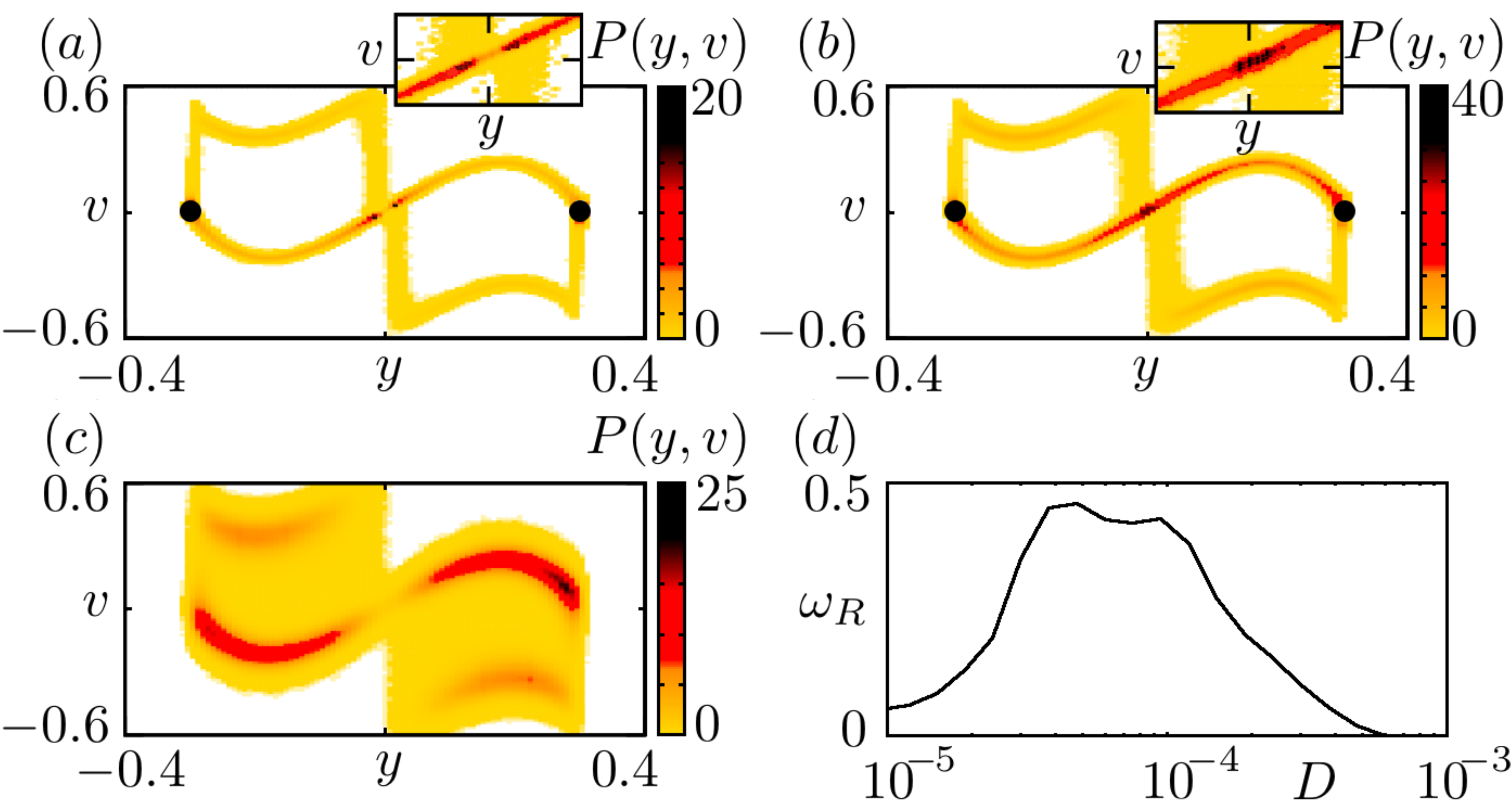} 
\caption{Excitable regime in analog simulations on example of the system's (\ref{system2}) electronic model. (a)-(c) Evolution of the PDF when the noise intensity increases: (a) $D=1.2\times 10^{-5}$; (b) $D=2.4\times 10^{-5}$; (c) $D=3.78\times 10^{-4}$. (d) Rice frequency, $\omega_{R}$, vs noise intensity, D. Parameters are: 
$\varepsilon=0.01$, $c_{1}=1$, $c_{3}=9$, $c_{5}=22$, $a=2.4$, $b=50$.}
\label{fig6}
\end{figure}  
%

As in the previous regime, there is difference between the behavior of the experimental facility and the dynamics of the mathematical model. In the presence of weak noise two closed craters in the experimentally obtained PDF are formed [Fig.~\ref{fig6}(a)]. Each crater has two maxima. The first maximum is situated near the origin and the second one corresponds to the stable equilibrium. In contrast to the PDF obtained in numerical simulation [Fig.~\ref{fig5}(a)], two peaks of either crater have very different heights. Therefore the maximal value $P(y,v)$ in two experimentally obtained PDF's [Fig.~\ref{fig6}(a,b)] is reduced in order to show the PDF transformations more evident. Thus, the peaks corresponding to the stable equilibria (are marked by black filled circles in Fig.~\ref{fig6}(a,b)) are truncated. Noise intensity growth initially leads to merge of two peaks near the origin [Fig.~\ref{fig6}(b)]. Then the peak in the origin splits into two peaks again if the noise growth continues, and the closed craters become separated again. Two craters become destroyed  [Fig.~\ref{fig6}(c)] and residuary peaks move away from the origin. Then transitions between left and right branches of the phase space become rare, and it requires extremely long experimental time realizations for the stationary PDF calculations. The experimentally obtained dependence of the Rice frequency on the noise intensity has two maxima [Fig.~\ref{fig6}(d)]. The existence of the first local maximum in the $\omega_{R}(D)$ dependence is associated with formation and further splitting of the peak of the PDF in the origin. Further noise-induced transitions do not occur when the noise intensity growths. Therefore, occurrence of the second local maximum in the $\omega_{R}(D)$ dependence is not a result of the noise-induced transitions, but is caused by an imperfection of the experimental setup.

\section{Conclusions} 
A generic model of the bistable oscillator with nonlinear dissipation has been studied in the self-oscillatory and excitable regimes. In the excitable regime the effect of coherence resonance is accompanied with multiple noise-induced transitions. The noise-induced transitions, registered as qualitative changes in the stationary PDF, were shown using analog circuit experiment and numerical simulation. Evolution of the power spectrum exactly conform to the transformation of the PDF. Thus, the noise intensity is an independent control parameter, which determines qualitative nature of the dynamics. Explanation of the mechanism of the described effects is based on partition of system's (\ref{system2}) phase space by nullclines and manifolds of the saddle equilibrium. 

Offered in \cite{semenov2016} stochastic model demonstrates a variety of regimes depending upon the parameter values and the noise-intensity. The simplest noise-induced phenomenon in the system (\ref{system2}) is described in \cite{semenov2016} and consists in transitions from the bimodal PDF to unimodal and back to bimodal when the noise intensity growths. This effect can be interpreted as noise-induced destruction and revival of the bistability. In the self-oscillatory and excitable regimes explored in the present paper one also can say about destruction and recovery of the bistability because of the qualitative changes in the PDF in the vicinity of the saddle equilibrium. In the presence of weak noise in both two regimes one can distinguish two separated closed craters. This shape of the PDF corresponds to hopping between two coexisting attractors and can be considered as a manifestation of the bistability. If the noise intensity growths, then the maximum of the PDF in the origin is formed. This peak is a common part of two craters, which were initially separated. It means that the separating boundary vanishes. In the presence of larger noise the PDF consists of two separated figures (craters or peaks) again. It is a typical PDF structure for the coexistence of two attractors (limit cycles or steady states) and noisy hopping between them. In that way one can say about the bistable behavior again. 

There are well-known studies of the simultaneous manifestation of coherence resonance and noise-induced transitions (see for example \cite{zakharova2010,geffert2014,semenov2015}), which are investigated as a correlated effects. In the present paper the coherence resonance and 
the noise induced transitions were considered as independent effects with different causes.

\section*{Acknowledgements} 
This work was supported by DFG in the framework of SFB 910, by the Russian Foundation for Basic Research (Grant No. 15-02-02288) and by the Russian Ministry of Education and Science (project code 3.8616.2017). I am very grateful to A.B. Neiman and T.E. Vadivasova and G.I. Strelkova for helpful discussions.

%

\end{document}